\documentclass{article}
\usepackage[utf8]{inputenc}
\usepackage{graphicx}
\usepackage[portrait, margin=0.8in]{geometry}
\usepackage{hyperref}
\usepackage{authblk}
\usepackage{todonotes}
\usepackage{enumitem}
\usepackage{float}
\usepackage{siunitx}
\usepackage{amsmath,amsthm,amssymb}
\usepackage{braket}
\usepackage[sorting = none, backend = bibtex, style=numeric-comp]{biblatex}

\title{Biased Degenerate Ground-State Sampling of Small Ising Models with Converged QAOA}

\author[1]{Elijah Pelofske\thanks{Email: epelofske@lanl.gov}}
\affil[1]{Los Alamos National Laboratory}

\date{\vspace{-6ex}}

\begin{document}
\maketitle

\begin{abstract}
The Quantum Alternating Operator Ansatz, a generalization of the Quantum Approximate Optimization Algorithm (QAOA), is a quantum algorithm used for approximately solving combinatorial optimization problems. QAOA typically uses the Transverse field mixer as the driving Hamiltonian. One of the interesting properties of the Transverse-field driving Hamiltonian is that it results in non-uniform sampling of degenerate ground states of optimization problems. 
In this study we numerically examine the fair sampling properties transverse field mixer QAOA, and Grover Mixer QAOA (GM-QAOA) which provides theoretical guarantees of fair sampling of degenerate optimal solutions, up to large enough $p$ such that the mean expectation value converges to an optimal approximation ratio of $1$. This comparison is performed with high quality heuristically computed, but not necessarily optimal, QAOA angles which give strictly monotonically improving solution quality as p increases. These angles are computed using the Julia based numerical simulation software JuliQAOA. Fair sampling of degenerate ground-states is quantified using Shannon entropy of the ground-state amplitudes distribution. The fair sampling properties are reported on several quantum signature Hamiltonians from previous quantum annealing fair sampling studies. Small random fully connected spin glasses are shown which exhibit exponential suppression of some degenerate ground-states with transverse field mixer QAOA. The transverse field mixer QAOA simulations show that some problem instances clearly saturate the Shannon entropy of $0$ with a maximally biased distribution that occurs when the learning converges to an approximation ratio of $1$ while other problem instances never deviate from a maximum Shannon entropy (uniform distribution) at any $p$ step. 

\end{abstract}

\section{Introduction}
\label{section:introduction}

Approximate quantum optimization algorithms are being studied for their potential to perform advantageous computing for heuristic sampling of combinatorial optimization problems. The most prominent quantum algorithms of this type are quantum annealing (QA) \cite{Kadowaki_1998, farhi2000quantum, Finnila_1994, morita2008mathematical, santoro2002theory} and the Quantum Approximate Optimization Algorithm (QAOA) \cite{farhi2014quantum}, which was subsequently generalized to the Quantum Alternating Operator Ansatz \cite{Hadfield_2019}. Typically, both QA and QAOA are implemented (both on near term quantum computing hardware and numerically) using the Transverse-field driving Hamiltonian. A clear advantage of using the Transverse-field driving Hamiltonian is that it is easy to implement on near term noisy quantum computing hardware; in the case of QAOA it requires only one layer of single qubit rotations per QAOA step, whereas other driving Hamiltonians can require significantly more circuit complexity to implement. Here, circuit complexity meaning specifically the number and connectivity of quantum gate operations of the QAOA circuit when compiled to a quantum computer architecture with a small number of basis gates. GM-QAOA requires higher circuit complexity to implement the mixing operation compared to transverse field mixer\cite{Bartschi_2020, Pelofske_2021, Golden_2022}, which uses only a single layer of single qubit rotations per $p$. The transverse field driving Hamiltonian used in these quantum algorithms has a clear signature of not sampling degenerate ground states fairly for all problem instances - as opposed to many classical algorithm analogues such as simulated annealing. Degenerate ground states are variable assignments which give the same globally optimal objective function cost value to the given optimization problem, but have potentially very different variable states. Generally the idea of fair sampling is motivated by low temperature sampling of Ising spin glass problems, equivalently combinatorial optimization problems, and generally it is the case that classical heuristic sampling algorithms such as simulated annealing will uniformly sample degenerate ground states (e.g., the distribution of optimal solution sampling has maximum Shannon entropy) \cite{PhysRevE.92.013303, MORENO_2003, kumar2020achieving}. This property of unfair optimal solution sampling of particular optimization problems by quantum approximate optimization has been used to validate the sampling properties of near-term D-Wave quantum annealing processors which use the transverse field driving Hamiltonian \cite{Boixo_2013, PhysRevA.91.042314, job2017testdriving, Albash_2015, shin2014quantum, smolin2013classical}.

While not always required for combinatorial optimization problems or general sampling problems, obtaining many (or all) of the degenerate optimal solutions, if there exist degenerate optimal solutions, is an important computational property for many sampling problem types including maximum clique numeration \cite{eblen2012maximum, chang2020efficient, 9250607, carchiolo2024geometric, 9477866, grassia2019learning}, the knapsack solution counting problem \cite{6108252}, SAT-based membership filters \cite{weaver2012satisfiability, douglass2015constructing, Azinovi__2017, fang2018naesatbased}, and propositional model counting \cite{gomes2009model, samer2010algorithms, sang2005heuristics, K_bler_2010, capelli2014hypergraph, vaezipoor2021learning}. This property is also quite relevant for physics simulations, including computing the structural properties or entropy of the ground-states of a degenerate spin system \cite{Zhu_2019, Sandvik_1999}, and Gibbs (or Boltzmann) sampling \cite{PhysRevApplied.17.044046, PRXQuantum.3.020317}. Therefore, it is important to consider how approximate quantum optimization will sample degenerate optimal solutions of combinatorial optimization problems, or equivalently spin systems. There have been several studies examining how biased quantum annealing sampling can be (both numerically and on noisy quantum hardware) \cite{PhysRevLett.118.070502, K_nz_2019, kumar2020achieving, Pelofske_2021, PhysRevA.97.022312}, potential mechanisms to mitigate this biased QA ground state sampling \cite{Zhu_2019, kumar2020achieving, Yamamoto_2020, PhysRevE.99.043306, PhysRevA.96.042322}, and even potentially cases where this sampling bias is useful for some computations \cite{Zhang_2017}. Generally, we expect such sampling bias to become relevant for relatively large scale implementations of QAOA being used as a sampler in these various contexts, such as a heuristic solver for combinatorial optimization. In regards to approximate quantum optimization with QAOA, there exist higher-order mixers that provide theoretical guarantees of fairly sampling states that share the same cost value. In particular Grover Mixer QAOA (GM-QAOA) \cite{Bartschi_2020}, which is based on the unstructured Grover Search algorithm \cite{grover1996fast}, always produces probability amplitudes for cost function values that are the same for all states sharing that cost value. However, numerical evidence for relatively small problem sizes has shown that GM-QAOA seems to require more steps, and therefore more computation time, compared to Transverse-field mixer QAOA \cite{Golden_2023_evidence, Golden_2023_SAT}. 

The sampling of degenerate ground-states by transverse field driving Hamiltonians is not always non-uniform - for example the trivial states of ferromagnetic or antiferromagnetic states are sampled fairly, and many random Ising models are also sampled fairly. However, there are test Ising models which do exhibit unfair degenerate ground state sampling in transverse magnetic field quantum annealing - and thus can be used as a probe to discern quantum annealing ground state sampling signature statistics.

Previous fair sampling QAOA studies have only examined up to $p=1$ (one round) QAOA circuits \cite{Pelofske_2021, Golden_2022}, $p=2$ \cite{zhang2024groverqaoa3satquadraticspeedup}, and $p=3$ \cite{Zhu_2022}. Therefore, it is a natural question of what effect, if any, does higher rounds of QAOA (e.g. beyond $p=1$) have on fair sampling - in particular for mixers with potentially biased ground state sampling such as the transverse field mixer. 

We also define, and evaluate, an ensemble of small random low coefficient precision ($\{+1, -1\}$) Sherrington-Kirkpatrick \cite{PhysRevLett.35.1792} models that both exhibit exponentially biased degenerate ground-state sampling when sampled using the driving transverse magnetic field in QAOA, as well as uniform sampling. We propose that these models could be used for probing of ground-state sampling statistics for other variants of QAOA and quantum annealing, as well as on NISQ computers. While not all types of combinatorial optimization problems exhibit this property (of biased degenerate ground-state sampling), it is important to know that this effect can occur - and could be a detrimental effect for some types of sampling problems, such as combinatorial optimization. Problems with high ground state degeneracy can exhibit biased sampling under some approximate quantum algorithms, but probing this property for real world, and therefore by definition quite large, optimization problems can be computationally challenging. This property of approximate quantum optimization algorithms will be increasingly important as larger scale optimization problems are solved using algorithms such as QAOA, and analysis of their fundamental algorithmic properties such as fair sampling will be increasingly relevant.

We will begin the discussion of approximate quantum optimization be defining QAOA. Any given version of QAOA is broadly defined by the following components:
\begin{itemize}[noitemsep]
    \item an initial state $\ket{\psi}$,
    \item a \texttt{phase separating} cost Hamiltonian $H_C$
    \item a \texttt{mixing} Hamiltonian $H_M$
    \item an integer $p\geq 1$, the number of rounds to run the algorithm \footnote{This parameter $p$ is referred to by a number of different terms in different contexts, including the number of \emph{layers}, the number of \emph{levels}, or QAOA \emph{depth}},
    \item two vectors of real numbers $\vec{\gamma} = (\gamma_1,...,\gamma_p)$ and $\vec{\beta} = (\beta_1,...,\beta_p)$, each with length $p$. These parameters are usually collectively called the \emph{QAOA angles}. 
\end{itemize}
The phase separator Hamiltonian adds different phases to states according to the cost function of the target optimization problem, in this way it encodes the cost values of the underlying combinatorial optimization problem $C(z)$ which has been formulated as a classical diagonal Hamiltonian. Then the mixer provides parameterized interference between solution vectors with different phases; the mixer induces transitions between classical states. QAOA consists of preparing this initial state $\ket{\psi}$, then performing alternating Hamiltonian simulations of the two non-commuting Hamiltonians $H_C$ and $H_M$ for $p$ steps, and at the end the state of all qubits are measured in the computational basis -- samples from this state are solutions to the original optimization problem $C(z)$. $H_C$ is parameterized by the size-independent parameters $\vec{\gamma}$, and $H_M$ is parameterized by the problem size-independent parameters $\vec{\beta}$. In this study, the standard QAOA objective value phase separator is used in all simulations. Two different mixers are used in this study; the standard transverse field mixer (also referred to as the X mixer, due to the Pauli X gate rotation), and the Grover Mixer \cite{Bartschi_2020}.

\section{Methods}
\label{section:methods}

The numerical simulations of Quantum Annealing, and the numerical simulations of angle high-quality finding and computing expectation values for different QAOA variants, were performed using the Julia \cite{bezanson2017julia} library and \texttt{JuliQAOA} \cite{Golden_2023}. The QAOA angle finding procedure is performed using $10,000$ basin hopping iterations, and $p$ is increased up until the point where the mean expectation value converges to the ground-state (optimal energy) up to numerical precision, which is set by the numerical floating point precision cutoff of $10^{-8}$. For combinatorial optimization this means that the ideal algorithm has converged to an approximation ratio of $1$. The entire learning process across the $10,000$ basin hopping iterations is not reported, instead we select the absolute best performing angles found during this learning process, and report ground state sampling statistics using those highly optimized angles. The \texttt{JuliQAOA} simulation package \cite{Golden_2023} has been utilized in several previous studies \cite{Golden_2023_evidence, Golden_2023_SAT, Pelofske_2023_QAOA, pelofske2023scalingwholechipqaoahigherorder, Rajakumar_2024}, and has been shown to generate extremely high quality QAOA angles (although, not necessarily optimal), thus removing the challenge of learning good QAOA angles at high rounds using standard machine learning approaches which can yield poor-performing angles. All numerical simulations are performed using full precision state vector simulations of \texttt{JuliQAOA} \cite{Golden_2023}, meaning that there is no finite sampling effect (e.g., shot noise). 

Solution quality will be quantified using both the energy of the Ising model and approximation ratio. The approximation ratio for a single vector of spins $z$ which has energy $z_e$ of an Ising model with a maximum energy of $C_{max}$ and a minimum energy of $C_{min}$ (across all $2^n$ variable states) is defined as $\frac{C_{max} - z_e}{C_{max} - C_{min}}$, where an approximation ratio of $1$ denotes the optimal variable assignment was found.

The optimization problem instances that are used as test cases first are a selection of small Ising models from previous fair sampling studies \cite{K_nz_2019, Matsuda_2009, Matsuda_2009_2} which have been used in previous fair sampling analysis studies \cite{Golden_2022, Pelofske_2021}, and random fully connected Ising models. Note that there are several different studies that have used different Ising models that have this biased ground state sampling characteristic - these are sometimes known as ``Quantum Signature Hamiltonians'' \cite{Boixo_2013, PhysRevA.91.042314, job2017testdriving, Albash_2015, shin2014quantum, smolin2013classical, PhysRevA.91.062320, kadowaki2019experimental, shin2014comment, wang2013comment}. 

Next, an ensemble of $5$ and $6$ spin fully connected Ising models are studied to examine which, if any, exhibit any amount of biased degenerate ground-state sampling. These Ising models have discrete linear and quadratic terms, and the Hamiltonian, equivalently the cost function $C(z)$, can be written as

\begin{equation}
    {\mathcal H} = \sum_{ij} J_{ij} \sigma_i^z \sigma_j^z + \sum_i h_i \sigma_i^z
\end{equation}

where $J_{ij}$ are the quadratic spin interactions and $h_i$ are the local fields. All of the Ising models used in this study are solved as \emph{minimization} combinatorial optimization problems. The ${\mathcal H}$ function evaluation for a particular $z$ vector is known as the energy of that configuration.

\begin{figure}[th!]
    \centering
    \includegraphics[width=0.32\textwidth]{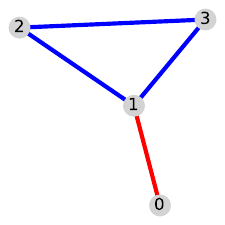}
    \includegraphics[width=0.32\textwidth]{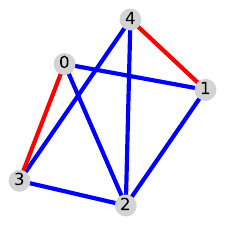}
    \includegraphics[width=0.32\textwidth]{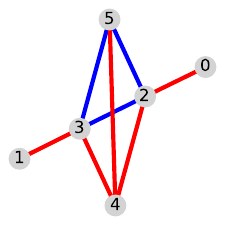}
    \includegraphics[width=0.32\textwidth]{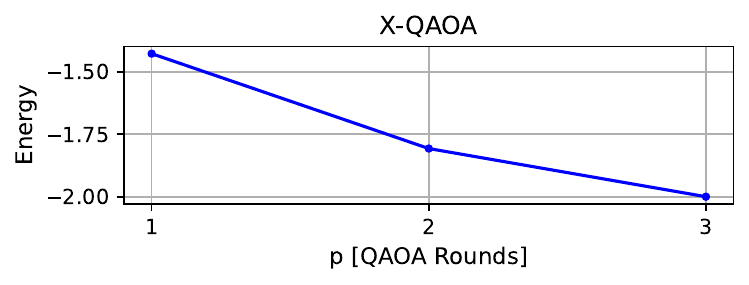}
    \includegraphics[width=0.32\textwidth]{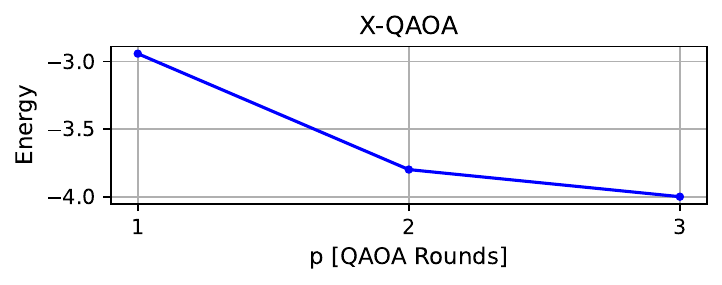}
    \includegraphics[width=0.32\textwidth]{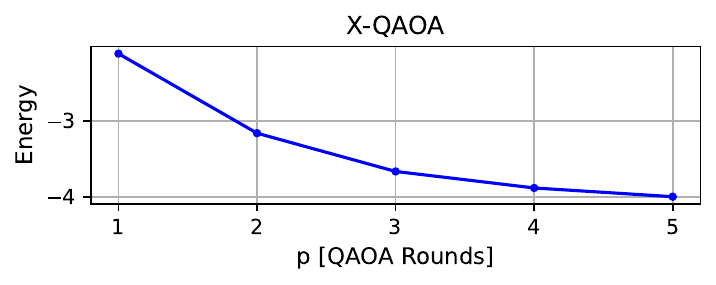}
    \includegraphics[width=0.32\textwidth]{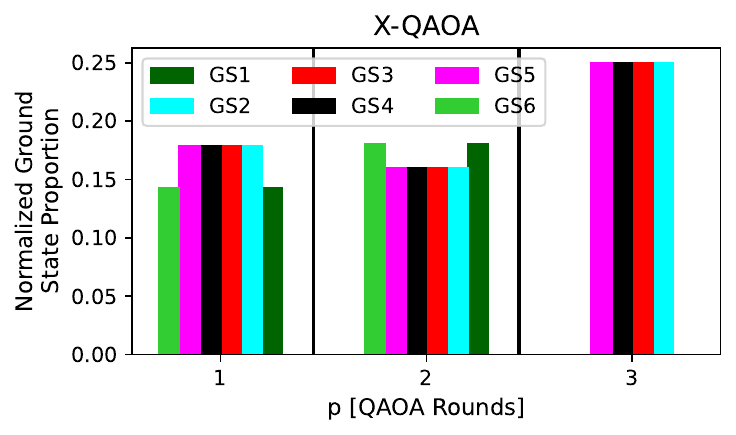}
    \includegraphics[width=0.32\textwidth]{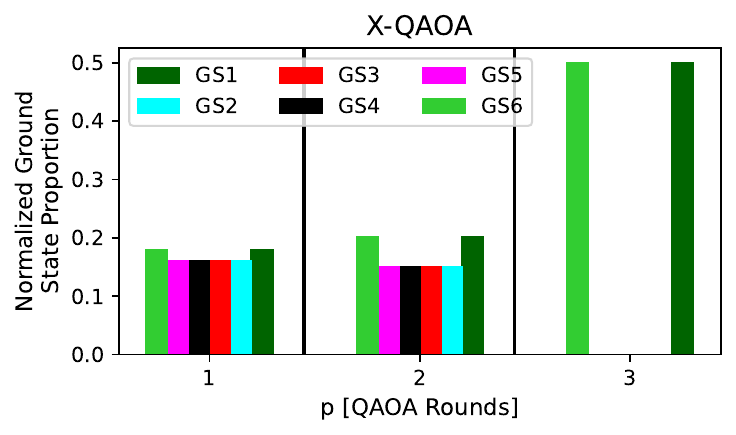}
    \includegraphics[width=0.32\textwidth]{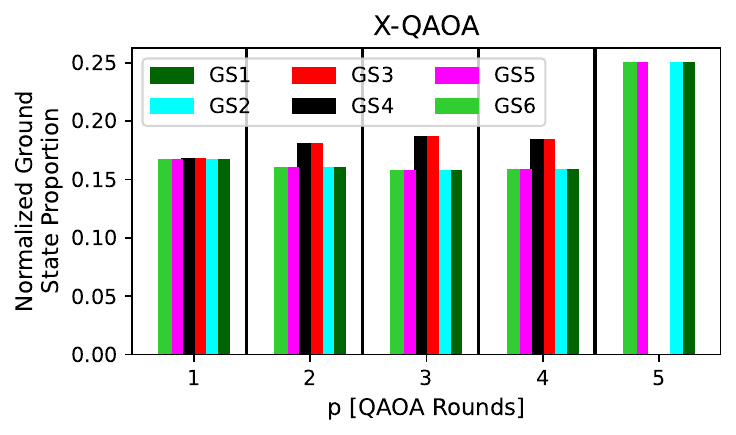}
    \includegraphics[width=0.32\textwidth]{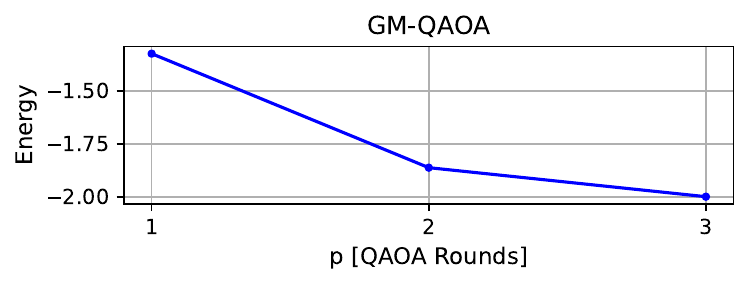}
    \includegraphics[width=0.32\textwidth]{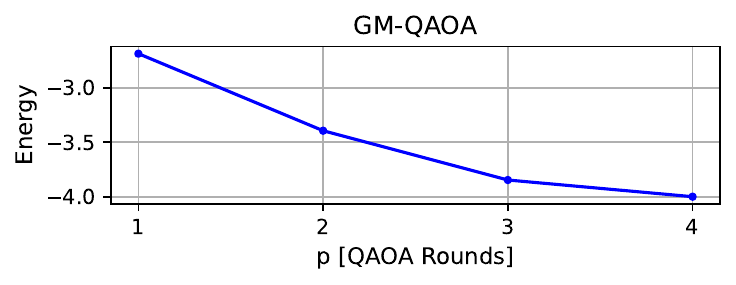}
    \includegraphics[width=0.32\textwidth]{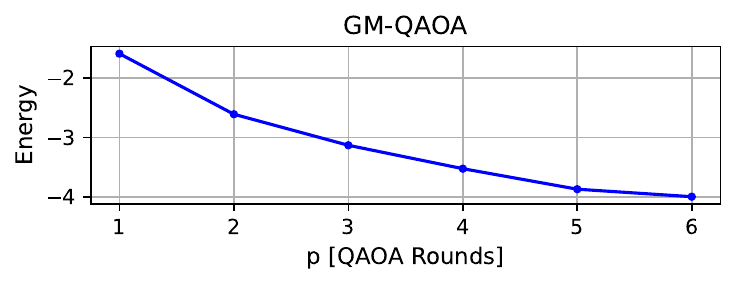}
    \includegraphics[width=0.32\textwidth]{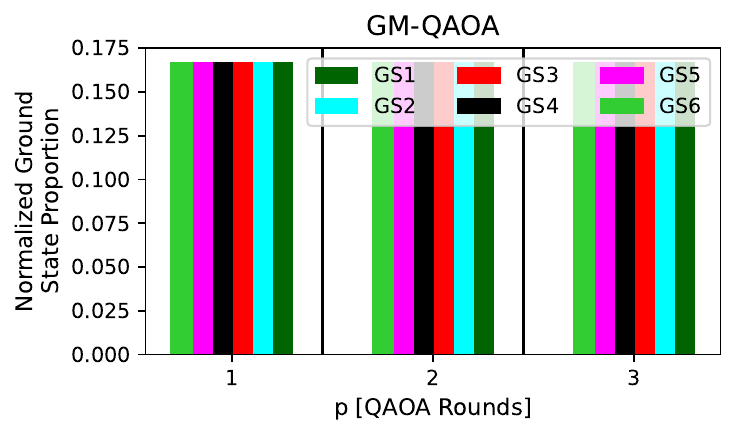}
    \includegraphics[width=0.32\textwidth]{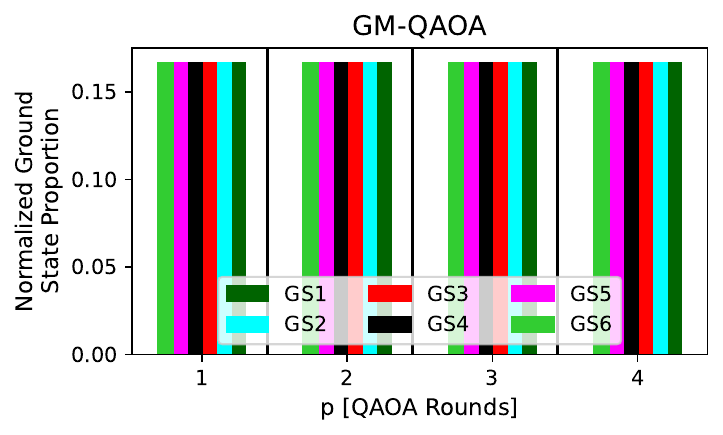}
    \includegraphics[width=0.32\textwidth]{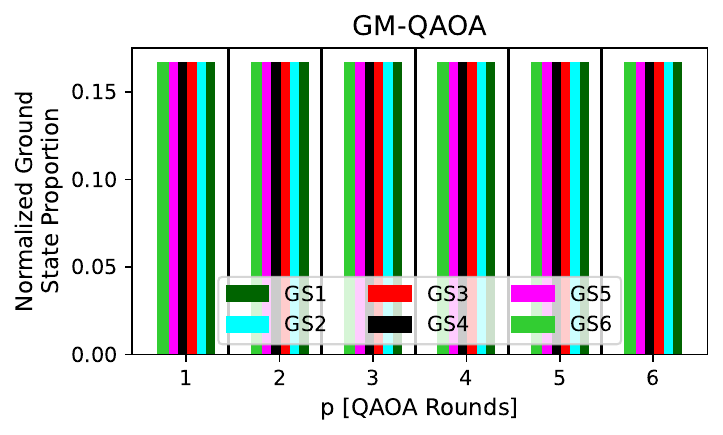}
    \caption{QAOA ground state fair sampling statistics on 3 test Ising models which have been shown in previous studies to have biased ground-state sampling properties when sampled with transverse field driven quantum annealing. The test Ising models are shown using blue to denote a coefficient of $-1$ (ferromagnetic), red to denote a coefficient of $+1$ (antiferromagnetic), and a light gray to denote a coefficient of $0$ (no local field). The plot is organized into columns where the test Ising model is at the top of the column, and then the X mixer QAOA (next two rows) and GM-QAOA ground state probabilities (last two rows), as well as energy performance curves, are shown below. The histograms show the exact ground state probability for a particular configuration, as a proportion out of the total ground-state probability, and each step of $p$ is delineated by vertical black lines. The energy sub-figures show the mean noiseless energy progression as a function of $p$, where each $p$ step strictly improves the energy due to the highly optimized QAOA angles. The X mixer QAOA probability histogram shows a clear bias, which becomes more pronounced as $p$ increases. }
    \label{fig:QAOA_fair_sampling_test_Ising_models}
\end{figure}

\begin{figure}[tbh!]
    \centering
    \includegraphics[width=0.44\textwidth]{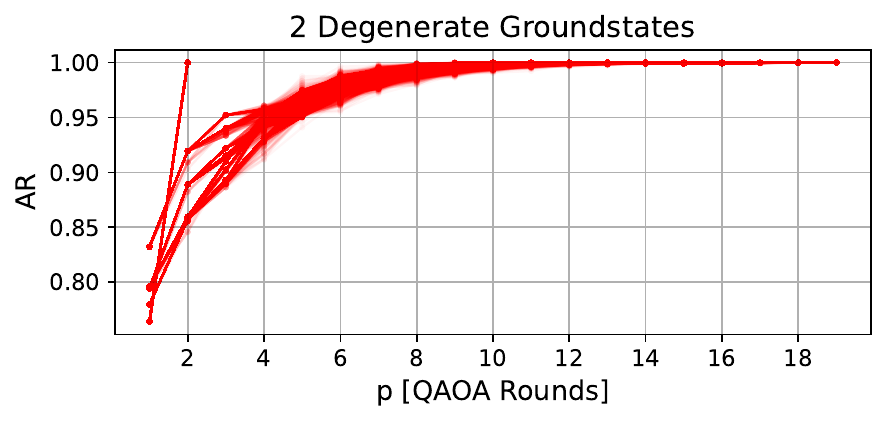}
    \includegraphics[width=0.44\textwidth]{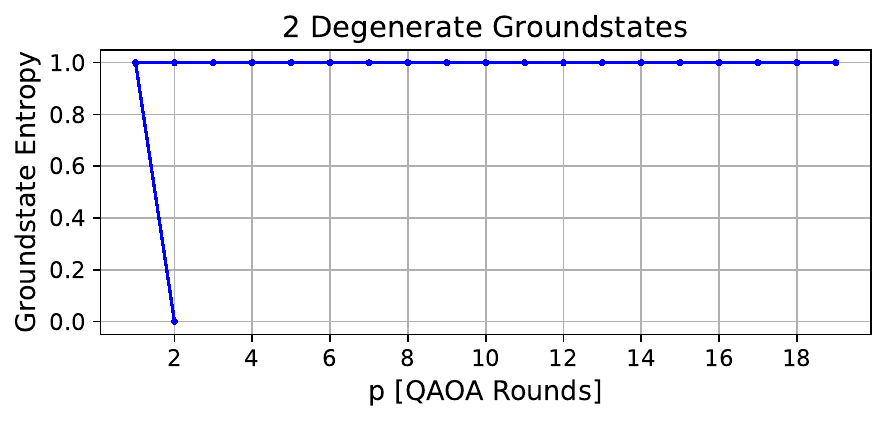}
    \includegraphics[width=0.44\textwidth]{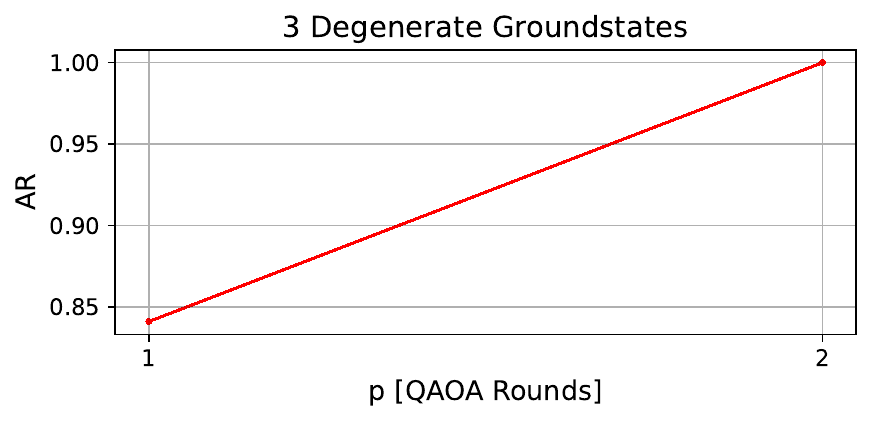}
    \includegraphics[width=0.44\textwidth]{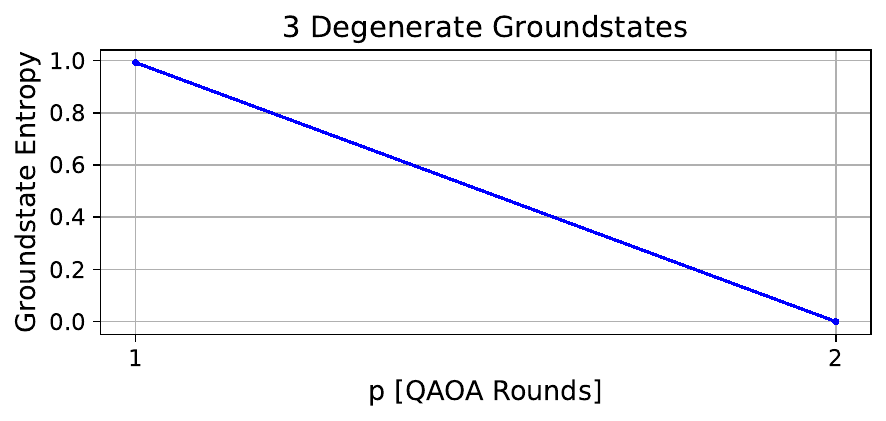}
    \includegraphics[width=0.44\textwidth]{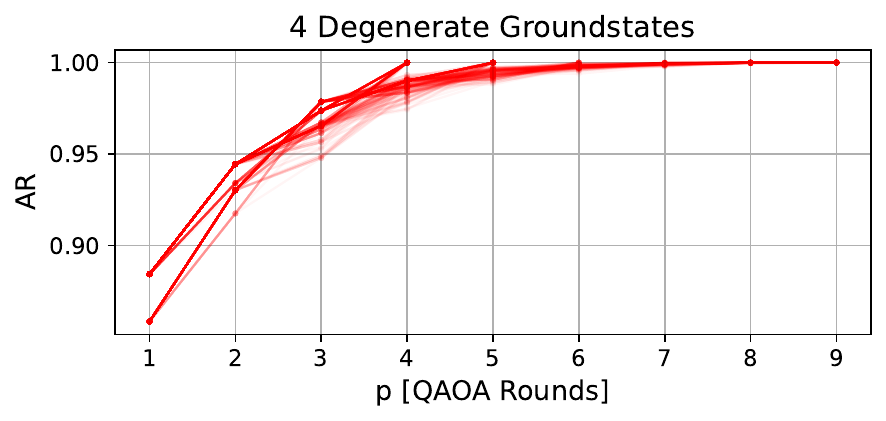}
    \includegraphics[width=0.44\textwidth]{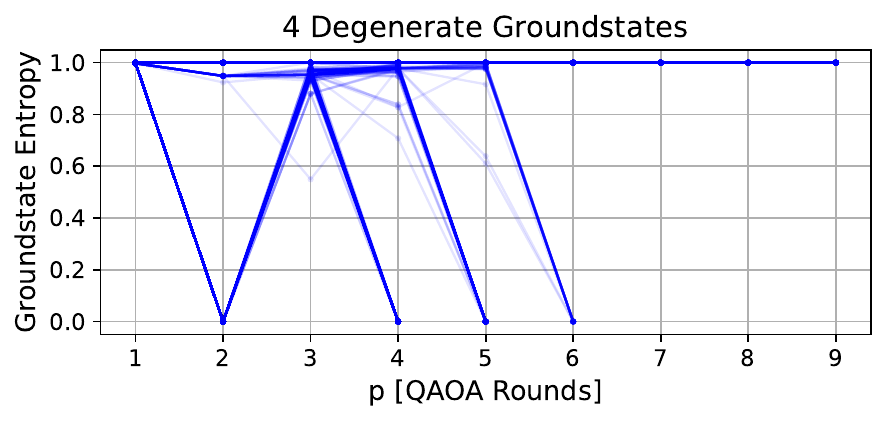}
    \includegraphics[width=0.44\textwidth]{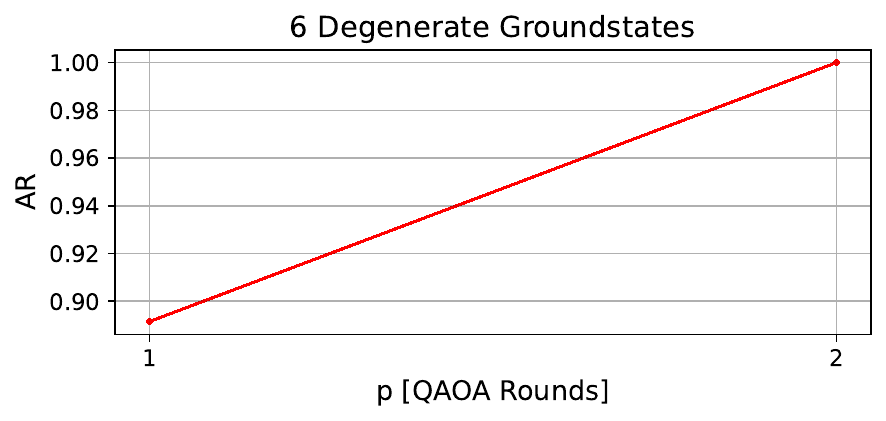}
    \includegraphics[width=0.44\textwidth]{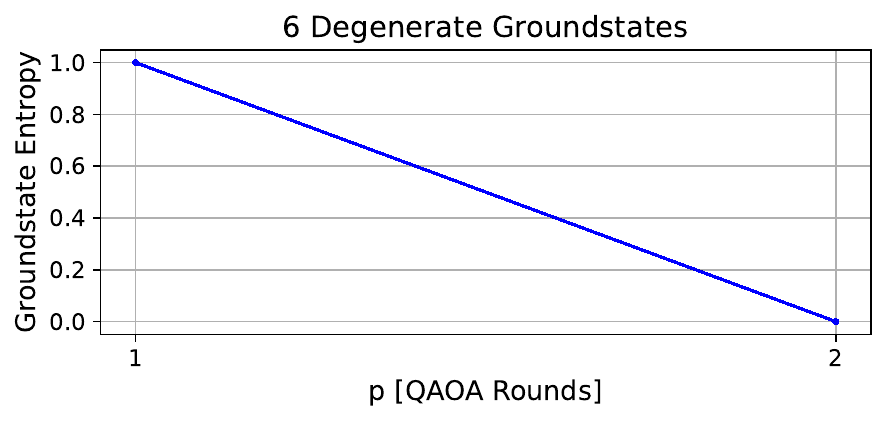}
    \includegraphics[width=0.44\textwidth]{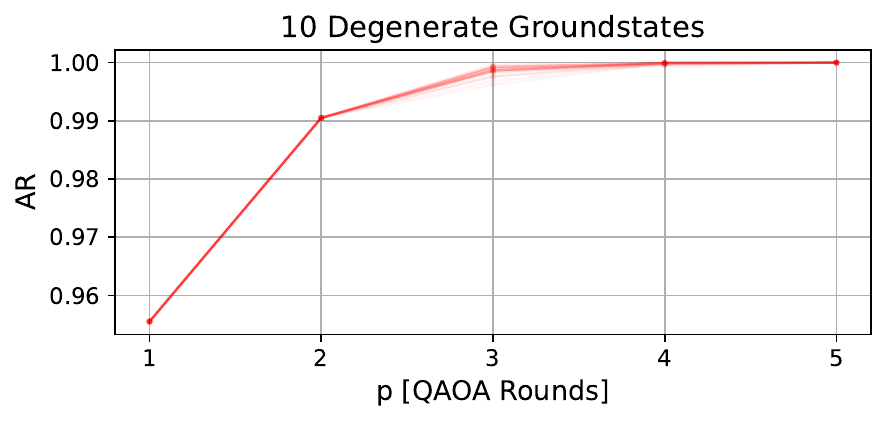}
    \includegraphics[width=0.45\textwidth]{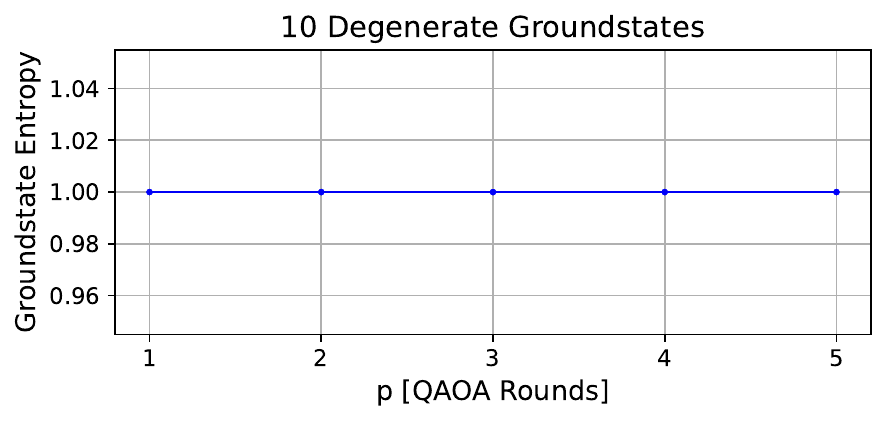}
    \caption{Transverse field mixer QAOA on random discrete-coefficient $5$ variable SK spin glass Ising model enumeration for different numbers of degenerate ground-states. The QAOA angle learning is terminated once the optimal energy converges to an approximation ratio (AR) of $1$, which can occur at different values of $p$ (left-column with red plot lines). The Shannon entropy of the distribution of ground-state probabilities (right-column with blue plot lines) quantifies how uniform or non-uniform the ground-state sampling is. The entropy is maximized ($1$) for uniform distributions and minimized ($0$) for the maximally non-uniform distribution (which is all of the ground-state probabilities contributing to a single degenerate ground-state spin vector). }
    \label{fig:SK_entropy}
\end{figure}

\section{Results}
\label{section:results}

Figure~\ref{fig:QAOA_fair_sampling_test_Ising_models} presents numerical simulations of the ideal (noiseless) sampling of the distinct spin variable configurations that have the same optimal energy, across three Ising models. Figure~\ref{fig:QAOA_fair_sampling_test_Ising_models} shows simulations from three of the test Ising models used in ref. \cite{K_nz_2019}. For these simulations, each sub-plot ends at the $p$ step where the mean energy has converged with very high probability to the optimal energy. This QAOA convergence comes due to the highly optimized angles that were found via numerical optimization, and here we are showing only the absolute best performing angles that were independently optimized for each $p$, for each Ising model, and for both mixers. With the transverse field mixer, the ground-state sampling is biased for all $p$ steps until convergence, whereas for GM-QAOA the optimal solutions are sampled uniformly as expected \cite{Bartschi_2020}. These three test Ising models are known to have biased ground-state sampling under transverse field driving Hamiltonian quantum annealing, and this property is shared by transverse field mixer QAOA. Notably, with the transverse field mixer the optimal solutions are not sampled uniformly at any $p$, however at the $p$ where the mean energy has fully converged to the optimal energy some of the degenerate spin configurations are exponentially suppressed to essentially $0$ probability. Another notable property is that the biased sampling has state symmetries (all of these optimal spin configurations are mirror-symmetric in the sense that spin flipping all of the variables also gives a ground-state) where spin symmetric states are sampled at the same probability (with can be either a majority of the measured probabilities, or be exponentially suppressed). For example, ground-state 1 and 6 (which are mirror-symmetric) are the states sampled the most for the second test Ising model at $p=3$, and the other $4$ ground-states are effectively not sampled (their probabilities are near or at $0$). The 5-spin model shown in Figure \ref{fig:QAOA_fair_sampling_test_Ising_models} shares very similar sampling characteristics to refs. \cite{K_nz_2019, Matsuda_2009, Matsuda_2009_2} where continuous time quantum annealing numerical simulations were performed of the same Ising model. 

In the particular case of these test case Ising models in Figure~\ref{fig:QAOA_fair_sampling_test_Ising_models}, we observe that the number of $p$ steps to reach convergence for GM-QAOA is larger than the transverse field mixer. This empirical finding has been observed in several previous studies \cite{Golden_2023_evidence, Golden_2023_SAT} - but it is not known whether this is an artifact of approximate angle finding, or if this is a fundamental computation difference between these two QAOA mixers. 

The learned angles that give these ground-state distributions in Figure~\ref{fig:QAOA_fair_sampling_test_Ising_models} are shown in Figures~\ref{fig:QAOA_angles_X_mixer} and \ref{fig:QAOA_angles_GM} in Appendix~\ref{section:appendix_QAOA_angles}.

\subsection{Degenerate Ground-State Analysis for Small Fully Connected Ising Models}
\label{section:results_example_Ising_models}

Figure~\ref{fig:SK_entropy} quantifies aggregated metrics for a large ensemble of small random discrete-coefficient SK models. The entropy and energy convergence sub-plots are separated based on the number of degenerate ground-states the given Ising model has. The motivation for this is because the fair sampling and energy convergence characteristics differ based on the number of degenerate ground-states the Ising model has; for example the Figure \ref{fig:SK_entropy} plots for $3$ and $6$ degenerate ground-states all converge to the optimal energy (an approximation ratio of $1$) in at most $p=2$ steps, whereas for the other Ising models convergence often took more $p$ rounds. The ensemble of Ising models was generated by enumerating through all $2^{15}$ coefficient configurations from $\{+1, -1\}$ for all linear and quadratic coefficients of the $5$ variable all-to-all Ising model. $24,192$ of the $32,768$ Ising models have only a single ground-state and therefore were not considered in the ground-state sampling fairness analysis. All of the remaining Ising models do have degenerate ground-states, which are given as follows: $7200$ of the Ising models have $2$ degenerate ground-states, $480$ of the Ising models have $3$ degenerate ground-states, $480$ of the Ising models have $4$ degenerate ground-states, $384$ of the Ising models have $6$ degenerate ground-states, and $32$ of the Ising models have $10$ degenerate ground-states. For all of those Ising models that have degenerate ground-states, we use \texttt{JuliQAOA} (with $20$ basin hopping iterations) to find improving transverse field mixer QAOA angles as a function of increasing $p$ up until the energy converges to an approximation ratio of $1$.

Figure~\ref{fig:SK_entropy} shows that whether the transverse field mixer QAOA suppresses some ground-state probabilities or not varies considerably across the different problem instances.

At each step of $p$ in Figure~\ref{fig:SK_entropy} the Shannon entropy of the distribution of the ground-state probabilities is measured - which is maximized for a uniform distribution of the ground-states. This entropy measure shows which of the Ising models were fairly sampled with transverse field mixer QAOA; if the entropy is consistently at $1$ for all $p$ then the ground-states are sampled fairly. Convergence of the entropy to $1$ is quantified by checking if all of the entropy measures for every step of $p$ is equal to $1$ (rounded to a precision of $8$ decimal places).

Of the Ising models shown in Figure~\ref{fig:SK_entropy} with $2$ degenerate ground-states, $5127$ were always sampled fairly (e.g., they have a ground-state entropy of $1$), meaning that $2073$ of the Ising models have a non-uniform ground-state distribution. The Ising models with $3$ and $6$ degenerate ground-states had a consistent non-uniform distribution. The Ising models that have $10$ degenerate ground-states have a consistent uniform ground-state distribution. Of the Ising models with $4$ degenerate ground-states, $160$ have a uniform ground-state distribution and therefore $320$ have a non-uniform distribution. Many of the non-uniform ground-state distributions can be seen in Figure~\ref{fig:SK_entropy} very clearly.

\begin{figure}[tbh!]
    \centering
    \includegraphics[width=0.24\textwidth]{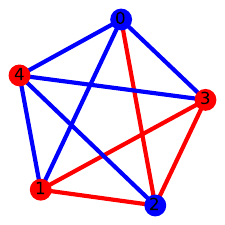}
    \includegraphics[width=0.24\textwidth]{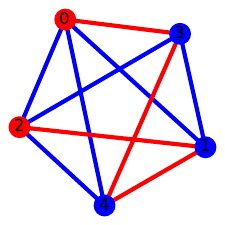}
    \includegraphics[width=0.24\textwidth]{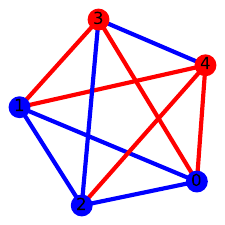}
    \includegraphics[width=0.24\textwidth]{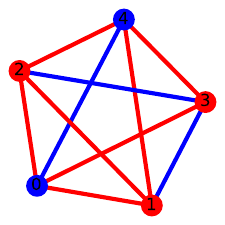}
    \includegraphics[width=0.24\textwidth]{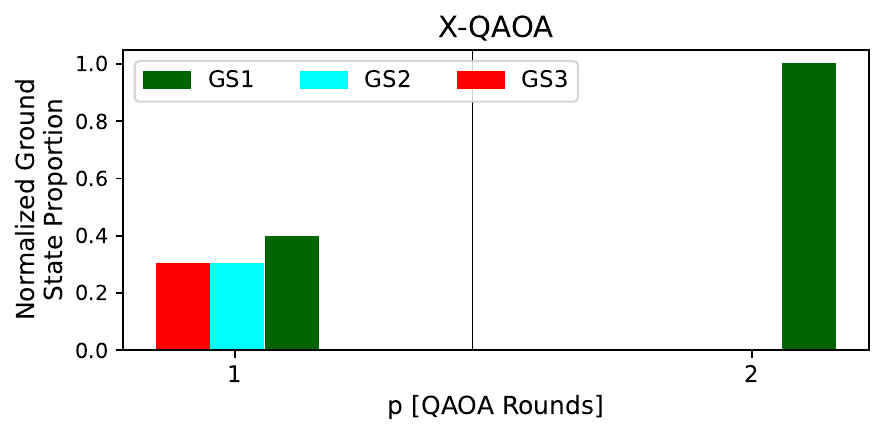}
    \includegraphics[width=0.24\textwidth]{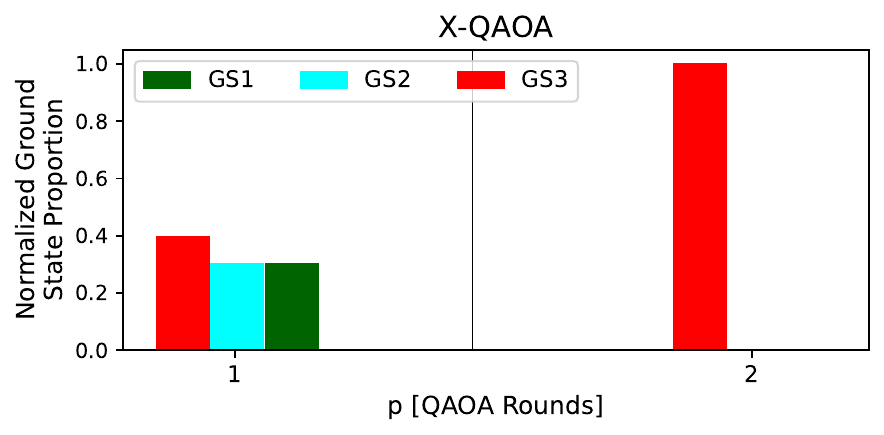}
    \includegraphics[width=0.24\textwidth]{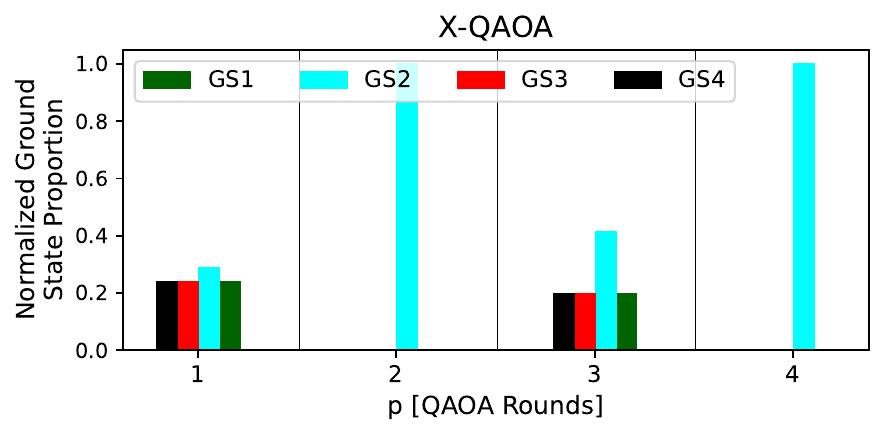}
    \includegraphics[width=0.24\textwidth]{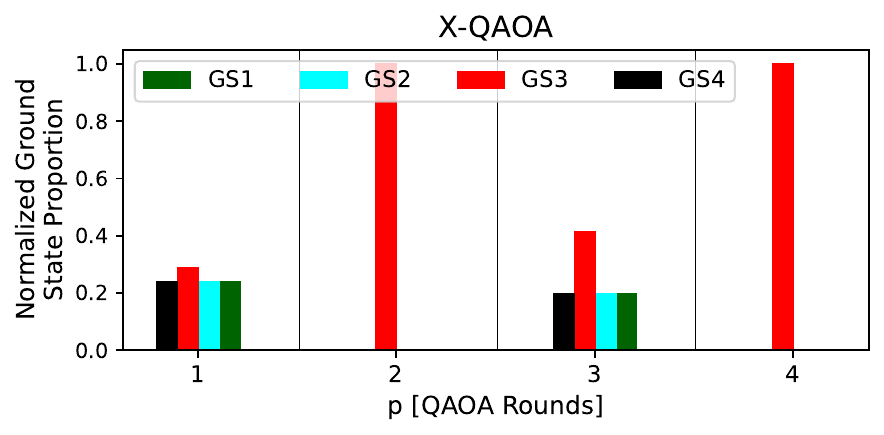}
    \includegraphics[width=0.24\textwidth]{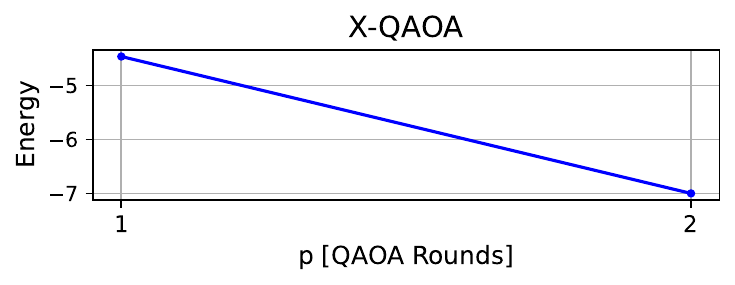}
    \includegraphics[width=0.24\textwidth]{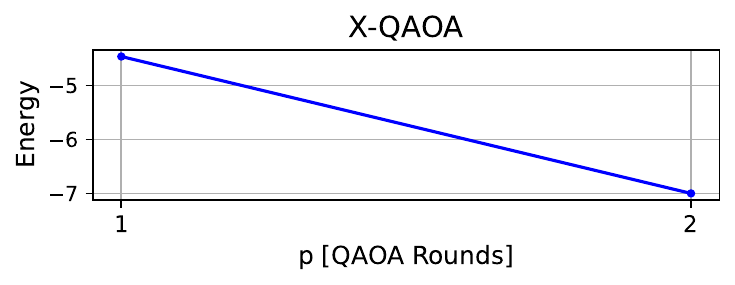}
    \includegraphics[width=0.24\textwidth]{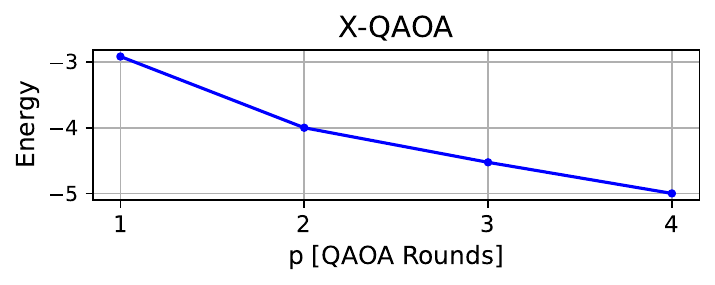}
    \includegraphics[width=0.24\textwidth]{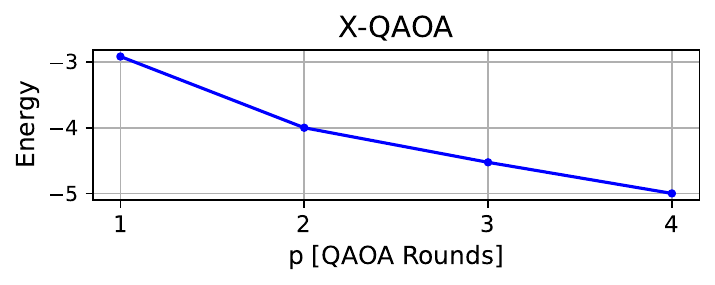}
    \includegraphics[width=0.24\textwidth]{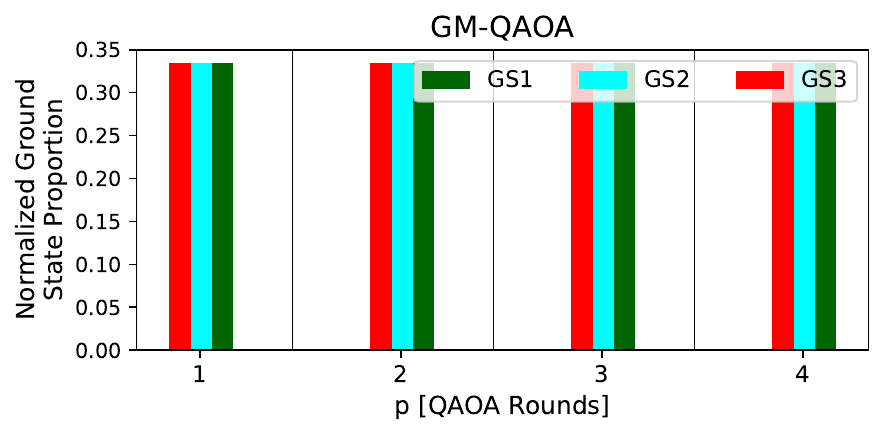}
    \includegraphics[width=0.24\textwidth]{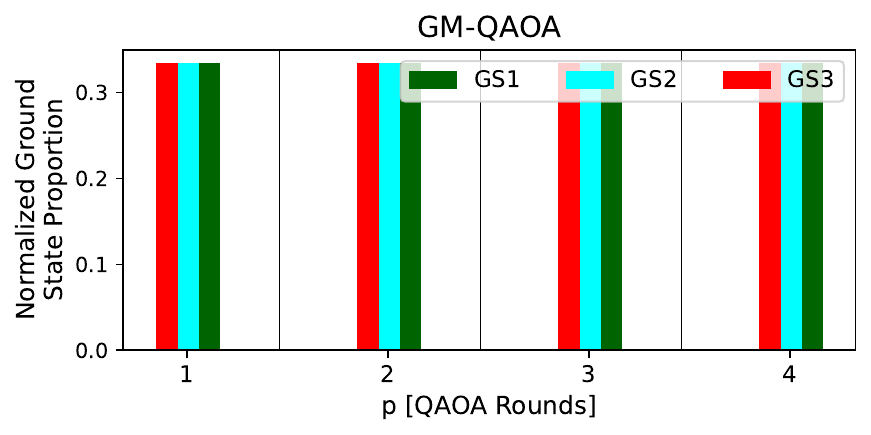}
    \includegraphics[width=0.24\textwidth]{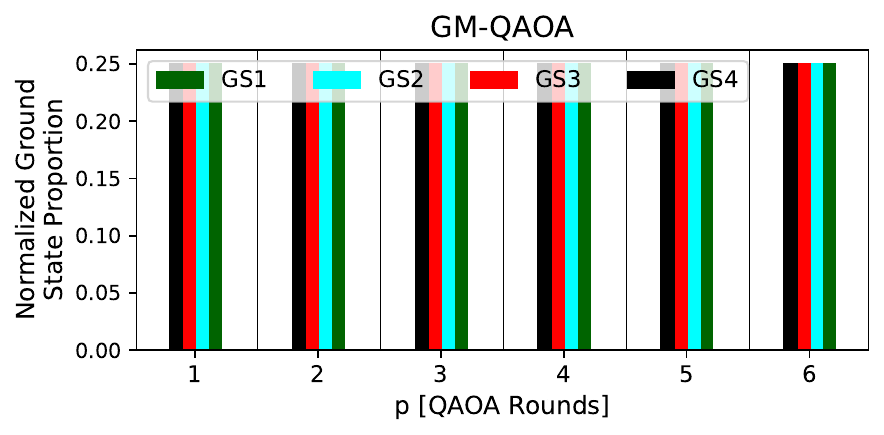}
    \includegraphics[width=0.24\textwidth]{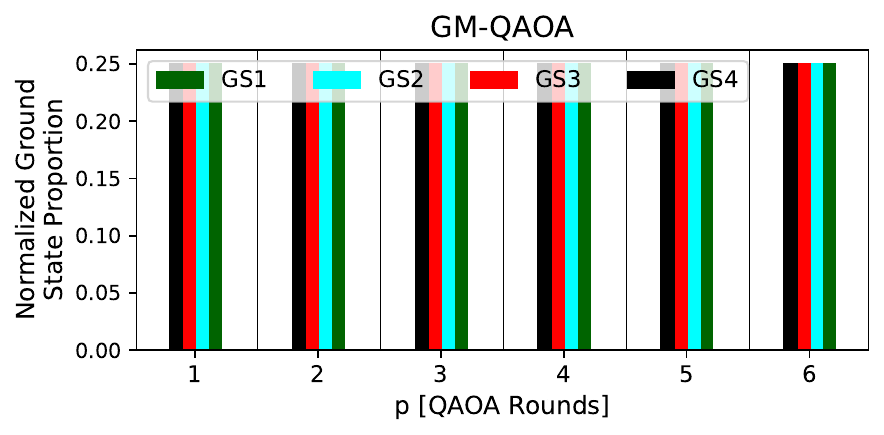}
    \includegraphics[width=0.24\textwidth]{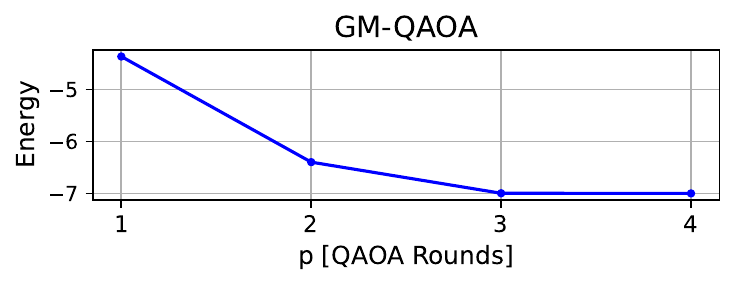}
    \includegraphics[width=0.24\textwidth]{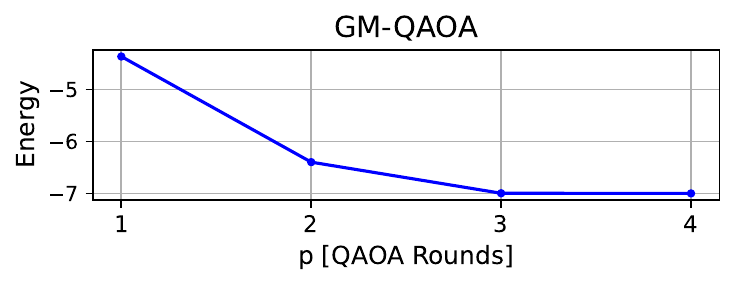}
    \includegraphics[width=0.24\textwidth]{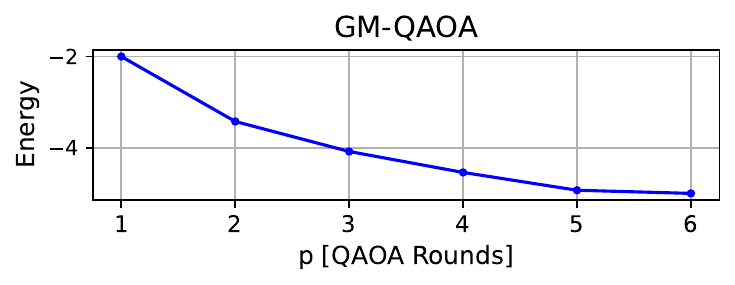}
    \includegraphics[width=0.24\textwidth]{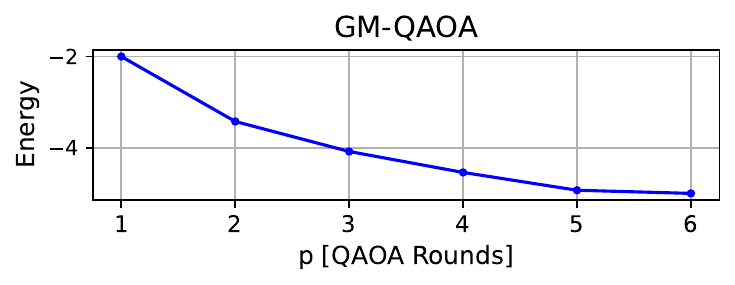}
    \caption{Four example Sherrington-Kirkpatrick models, each having exactly $5$ spins, when sampled with transverse field mixer QAOA exhibit biased ground-state sampling. This figure is organized into four columns, where each column corresponds to one of the example SK models that we found via enumeration that exhibit very clear ground state probability bias under the transverse field mixer. The Ising model is defined at the top of each column, and then the probability distributions for the highly optimized QAOA angles as well as the energy given by those same QAOA angles are then shown in the bottom four rows. Top row: graphical representation of the coefficient weights of each example Ising model where blue denotes a coefficient (either linear or quadratic) of $-1$, and red denotes a coefficient of $+1$. Middle two rows: histograms of the ground-state sampling proportion along with the corresponding energy found at that $p$ iteration (up until convergence to optimal energy) using the Transverse field mixer. Bottom two rows: same histogram and energy plots as before, but now using Grover Mixer QAOA. Notably, at the largest $p$, in all four cases the ground state probabilities are all concentrated into a single configuration. }
    \label{fig:example_SK_models_1}
\end{figure}

Figure~\ref{fig:SK_entropy} also shows that the convergence to an approximation ratio of $1$ can happen in as few as $p=2$ steps, or take up to $p=19$. Note that here convergence to the maximal approximation ratio of $1$ is determined by the numerical precision limit of $10^-8$ within the true ground-state energy, as described in Section~\ref{section:methods}. Importantly, all of the QAOA energy curves in Figure \ref{fig:SK_entropy} are strictly improving (since these Ising models are being solved as \emph{minimization} problems, this means that the mean energy is strictly decreasing as a function of $p$). This is not always trivial to compute, and in particular different runs of \texttt{JuliQAOA} are stochastic (because the $p=1$ angle choice is randomized) - what we found is that at reasonably high rounds, although the energies always converged, all steps of $p$ did not necessarily produce strictly improving energies. Therefore, we re-run the angle finding with \texttt{JuliQAOA} (with $20$ basin hopping iterations) on these cases until the energy is strictly improving for each discrete step of $p$.

\begin{figure}[tbh!]
    \centering
    \includegraphics[width=0.24\textwidth]{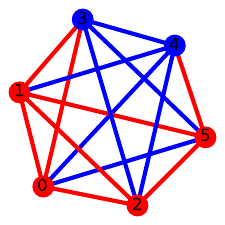}
    \includegraphics[width=0.24\textwidth]{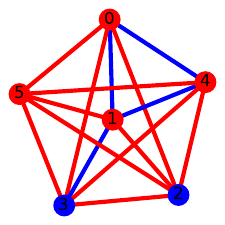}
    \includegraphics[width=0.24\textwidth]{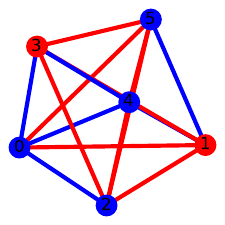}
    \includegraphics[width=0.24\textwidth]{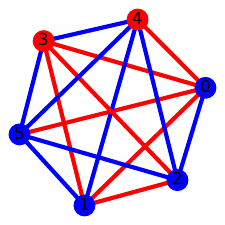}
    \includegraphics[width=0.24\textwidth]{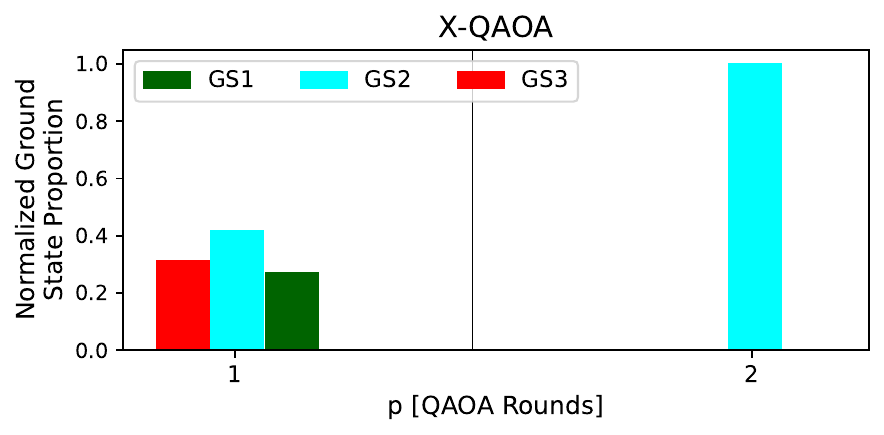}
    \includegraphics[width=0.24\textwidth]{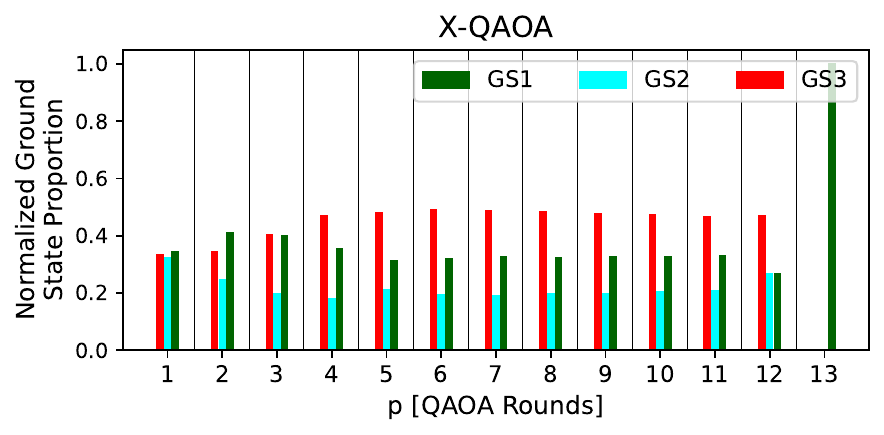}
    \includegraphics[width=0.24\textwidth]{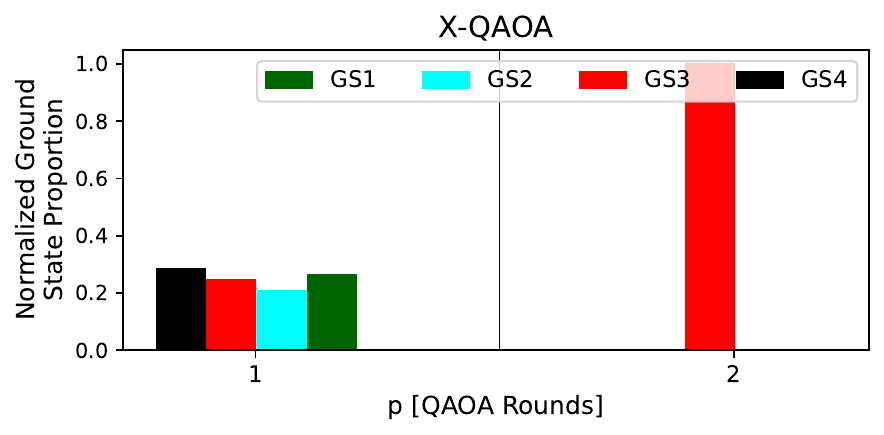}
    \includegraphics[width=0.24\textwidth]{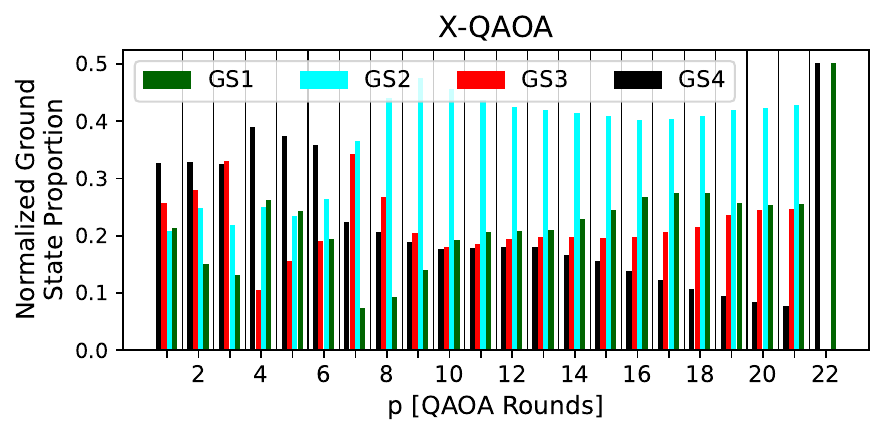}
    \includegraphics[width=0.24\textwidth]{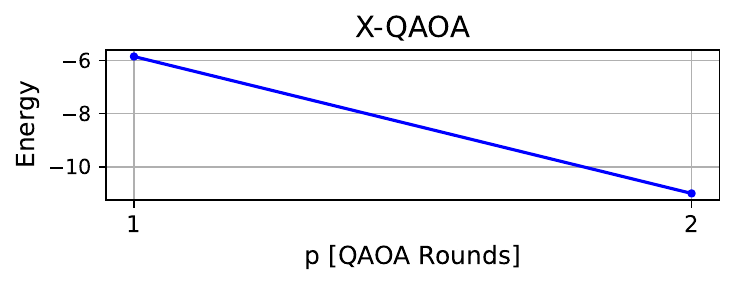}
    \includegraphics[width=0.24\textwidth]{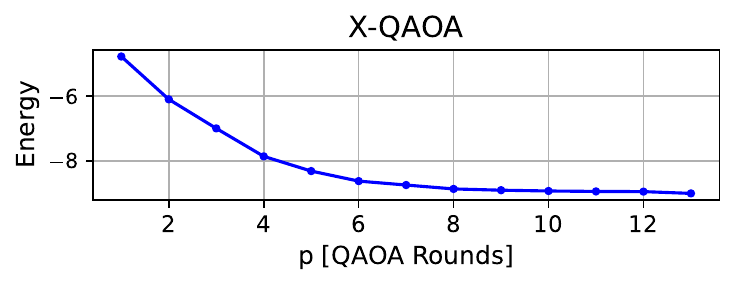}
    \includegraphics[width=0.24\textwidth]{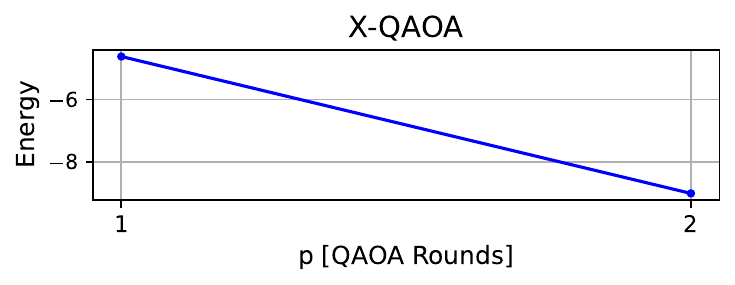}
    \includegraphics[width=0.24\textwidth]{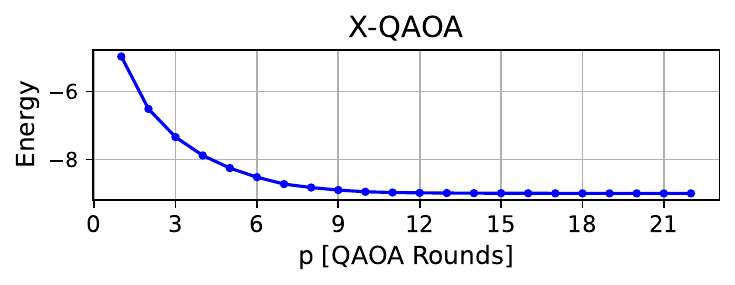}
    \includegraphics[width=0.24\textwidth]{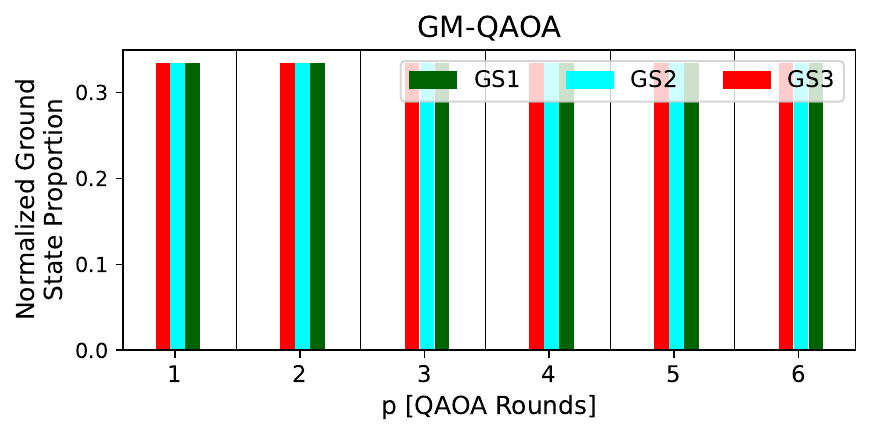}
    \includegraphics[width=0.24\textwidth]{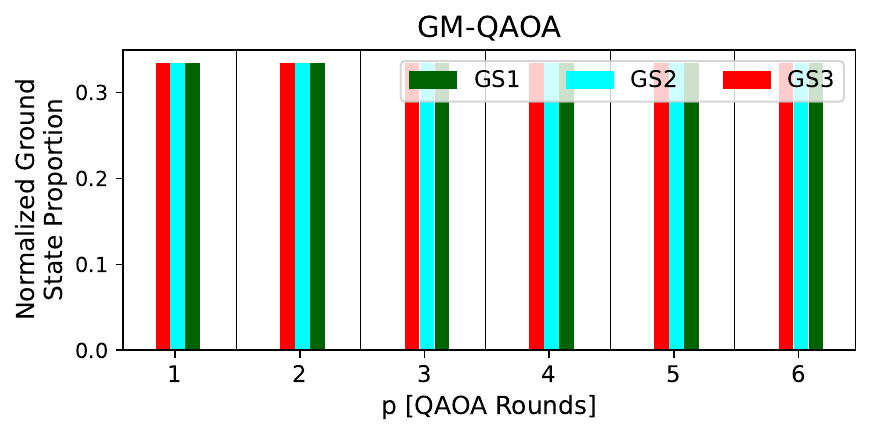}
    \includegraphics[width=0.24\textwidth]{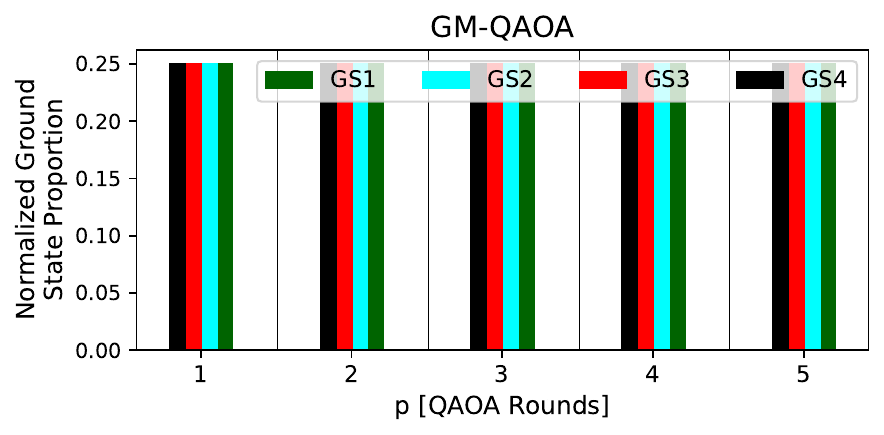}
    \includegraphics[width=0.24\textwidth]{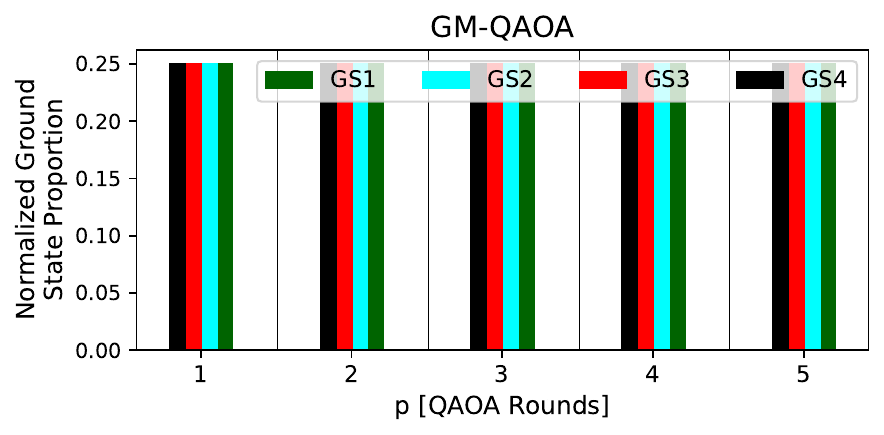}
    \includegraphics[width=0.24\textwidth]{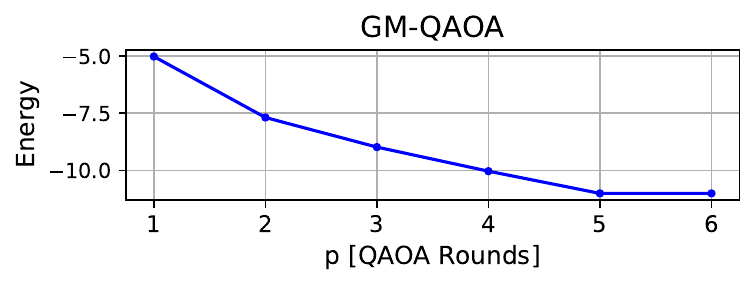}
    \includegraphics[width=0.24\textwidth]{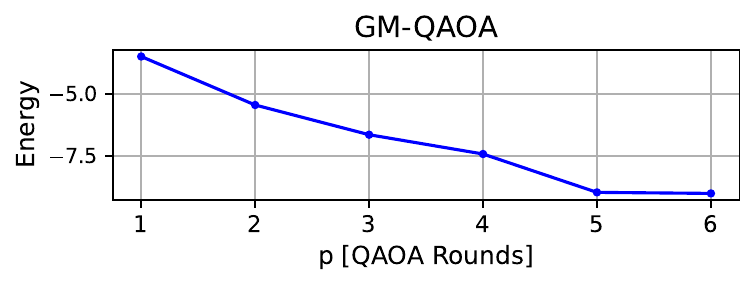}
    \includegraphics[width=0.24\textwidth]{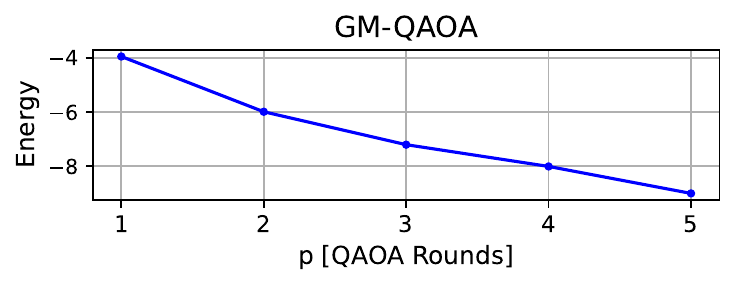}
    \includegraphics[width=0.24\textwidth]{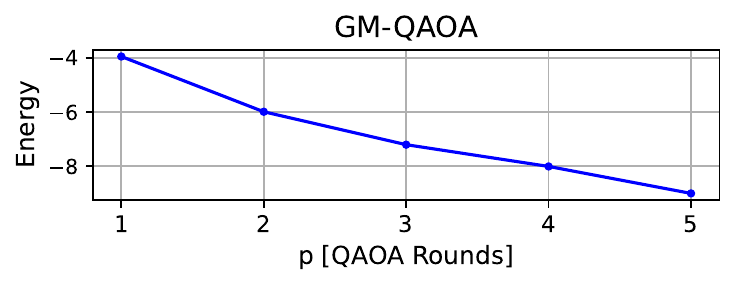}
    \caption{Four example Sherrington-Kirkpatrick models, each with $6$ spins, that when sampled with transverse field mixer QAOA exhibit biased ground-state sampling. This plot follows the same format as Figure~\ref{fig:example_SK_models_1}. Top row: graphical representation of the coefficient weights of each example Ising model, the QAOA simulations for which are shown in the column below each model. Middle two rows: histograms of the ground-state sampling proportion along with the corresponding energy found at that $p$ iteration (up until convergence to optimal energy) using the Transverse field mixer. Bottom two rows: same histogram and energy plots as before, but now using Grover Mixer QAOA. Notably, two of these Ising models require a larger number of $p$ steps (up to $p=22$) to converge, but once they are converged the probabilities concentrate into only one or two configurations.  }
    \label{fig:example_SK_models_2}
\end{figure}

Figures \ref{fig:example_SK_models_1} and \ref{fig:example_SK_models_2} combined show $8$ specific examples of $5$ and $6$ variable fully connected Ising models that exhibit clear non-uniform degenerate ground-state sampling when sampled using transverse field mixer QAOA. These example Ising models were hand-picked because they do exhibit biased ground-state sampling, although as Figure \ref{fig:SK_entropy} shows not all of these fully connected Ising models have biased ground-state sampling with transverse field mixer QAOA. The primary notable feature of the progression of ground-state sampling as $p$ increases is that, similar to Figure \ref{fig:QAOA_fair_sampling_test_Ising_models}, once QAOA has converged to an approximation ratio of $1$ some of the optimal spin configurations are heavily suppressed. All Ising model instances are rendered by showing Ising model coefficients encoded as blue to denote a coefficient of $-1$ (ferromagnetic) and red to denote a coefficient of $+1$ (antiferromagnetic).

\section{Discussion and Conclusion}
\label{section:conclusion}

Sampling spin glass Ising models that have multiple optimal solutions (e.g., ground-state degeneracy) using transverse mixer QAOA can, for some problems, results in suppression of some of the degenerate ground-states - this effect becomes more pronounced as the mean approximation ratio approaches $1$. GM-QAOA when applied to the same test Ising models gives strictly uniform ground-state sampling as expected. 

This study gives an analysis of small test case Ising models, whose sampling properties can be examined when being solved by approximate quantum algorithms, including specific instances that could be used to validate QAOA characteristics on near term quantum computers and can be easily validated using numerical QAOA simulations. 

Given the increasing interest in development and testing of quantum algorithms that can solve combinatorial optimization problems, in part due to the prevalence and important of being able to efficiently solve combinatorial optimization problems, it is important to understand the characteristics of how these algorithms work. Not being able to uniformly sample solutions to optimization problems that have the same cost values is a property which is not present in most standard classical sampling optimization algorithms. It is important to know that this property can exist in some quantum algorithms - and also that it does not occur in other quantum algorithms such as GM-QAOA - for future quantum algorithm development. 

However, this biased ground-state sampling property would need to be studied for larger system sizes to understand potential impacts on quantum algorithmic. This can be a challenging question to answer systematically because it requires knowing all degenerate ground-states which in the worst case requires brute force enumeration of all $2^n$ spin configurations. There is no reason to expect that this ground state biasing effect will diminish with larger system sizes, however there is a lack of systematic studies of degenerate ground state sampling at larger system sizes. 

One other aspect of the concept of fair sampling is the rate of convergence, e.g., solution quality, and how this is related to the uniformity of optimal configuration sampling. In this study, we focus primarily on the uniform sampling aspect of QAOA, and while the rate of convergence is certainly important algorithmically, here we aimed to focusing largely on the uniformity aspect. It seems, that there could exist a tradeoff between computational resources and uniformity of degenerate ground state sampling \cite{Pelofske_2021, Golden_2022, Golden_2023_evidence, Golden_2023_SAT}. 

This study has examined the fairness of optimal solution sampling, however the diversity of sampling sub-optimal solutions can also be examined \cite{PhysRevE.108.065303, zucca2021diversity} - in particular GM-QAOA fairly samples \emph{all} states based on cost function value, meaning it could sample rare solutions better than alternative approximate quantum optimization. Future study could investigate how measures such as Hamming distance between solutions, including non-optimal solutions, determines, if at all, how spin configurations are fairly or unfairly sampled. 

Future study could also examine the fair sampling properties of other QAOA mixers \cite{Hadfield_2019}, such as Hamming weight constrained mixers for example the $XY$ mixer \cite{Wang_2020}. There are an increasing number of variants of QAOA, not only limited to different mixing unitaries, but also ideas such as different initial states -- all of these variants should be examined in future study for their fair sampling properties. Fair sampling of these mixers would be applied to specific classes of optimization problems that naturally have a Hamming weight constraint.

\section{Acknowledgments}
\label{section:acknowledgments}
This work was supported by the U.S. Department of Energy through the Los Alamos National Laboratory. Los Alamos National Laboratory is operated by Triad National Security, LLC, for the National Nuclear Security Administration of U.S. Department of Energy (Contract No. 89233218CNA000001). Research presented in this article was supported by the NNSA's Advanced Simulation and Computing Beyond Moore's Law Program at Los Alamos National Laboratory. LANL report number LA-UR-24-30395.

\appendix

\section{QAOA Angles}
\label{section:appendix_QAOA_angles}

\begin{figure}[th!]
    \centering
    \includegraphics[width=0.32\textwidth]{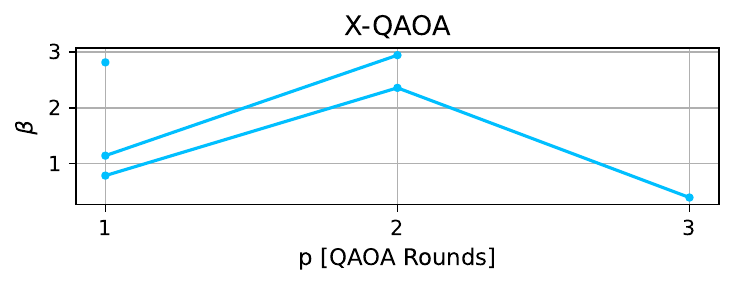}
    \includegraphics[width=0.32\textwidth]{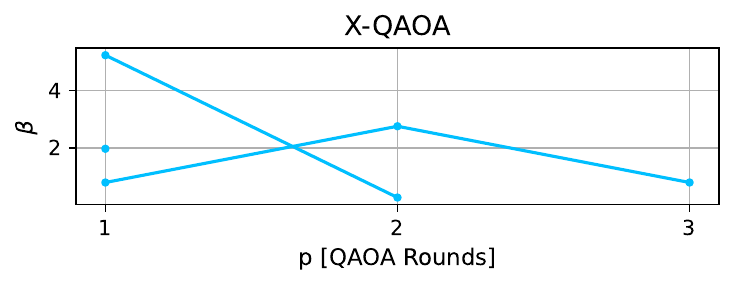}
    \includegraphics[width=0.32\textwidth]{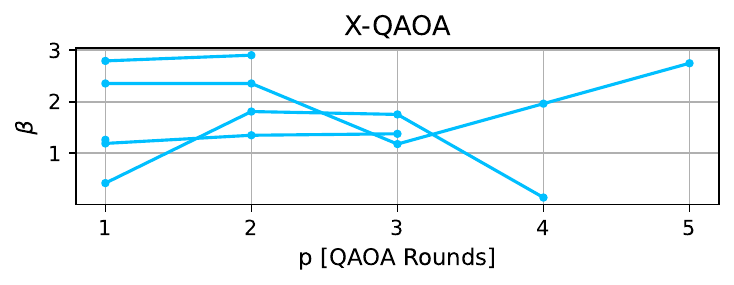}
    \includegraphics[width=0.32\textwidth]{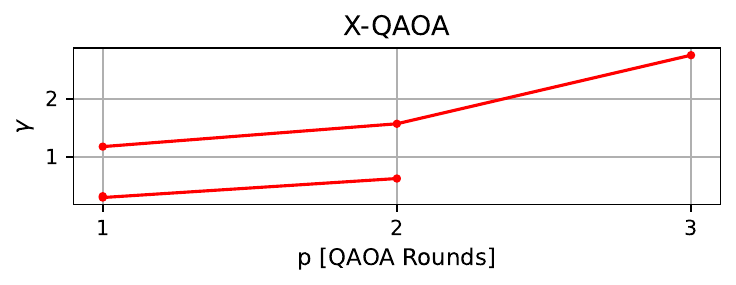}
    \includegraphics[width=0.32\textwidth]{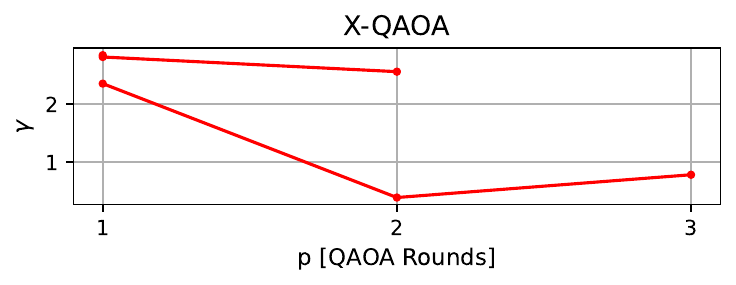}
    \includegraphics[width=0.32\textwidth]{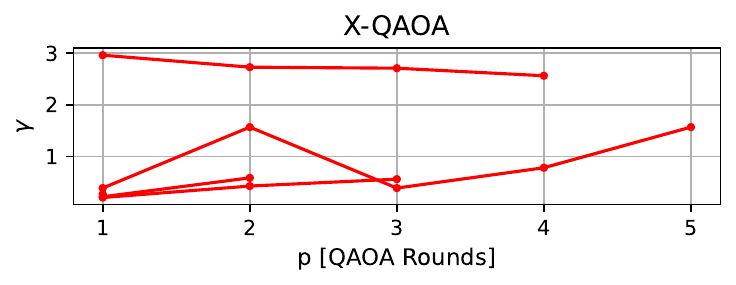}
    \caption{Learned QAOA angles from the Transverse field mixer QAOA simulations for the $3$ Ising models shown in Figure~\ref{fig:QAOA_fair_sampling_test_Ising_models}, up until convergence to the ground-state. The three vertical columns correspond to the three Ising models shown in Figure~\ref{fig:QAOA_fair_sampling_test_Ising_models}. For each step of $p$, the complete vectors of $\vec{\beta}$ (top row) and $\vec{\gamma}$ (bottom row) are shown.  }
    \label{fig:QAOA_angles_X_mixer}
\end{figure}

\begin{figure}[th!]
    \centering
    \includegraphics[width=0.32\textwidth]{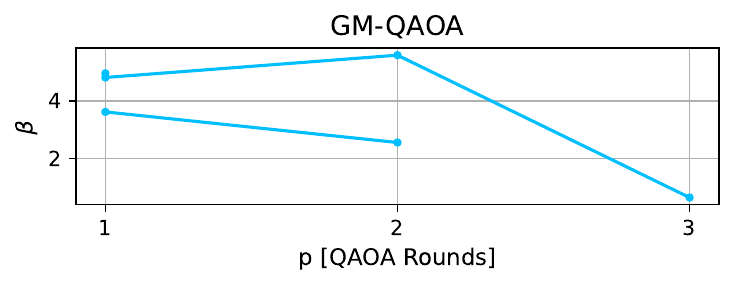}
    \includegraphics[width=0.32\textwidth]{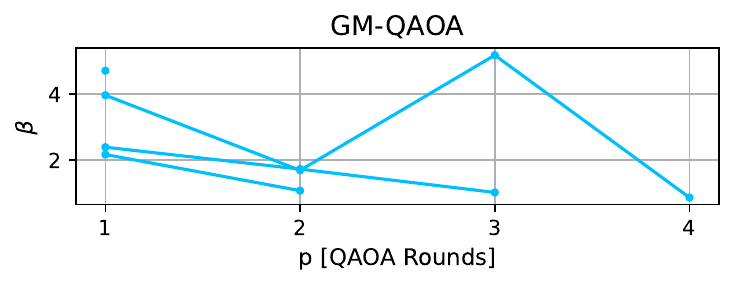}
    \includegraphics[width=0.32\textwidth]{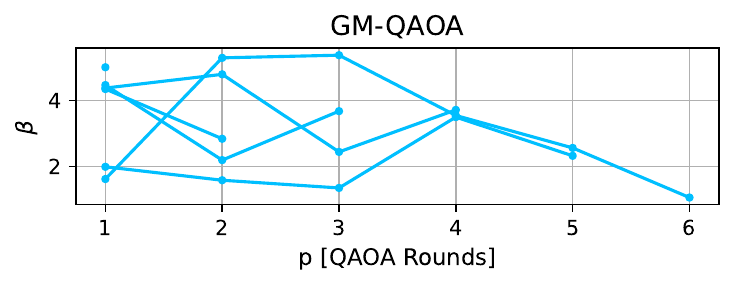}
    \includegraphics[width=0.32\textwidth]{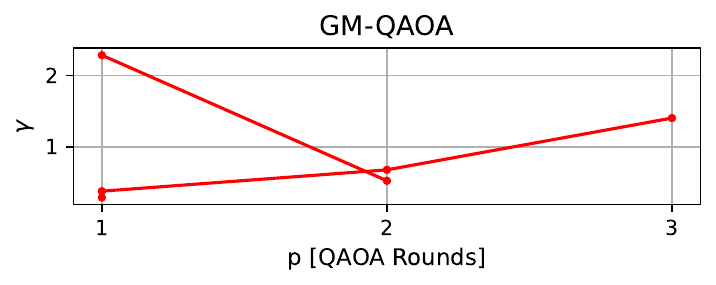}
    \includegraphics[width=0.32\textwidth]{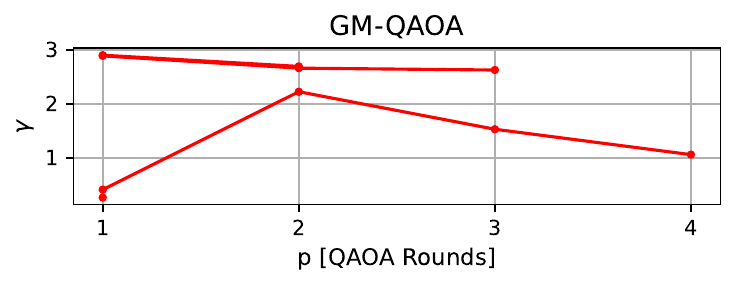}
    \includegraphics[width=0.32\textwidth]{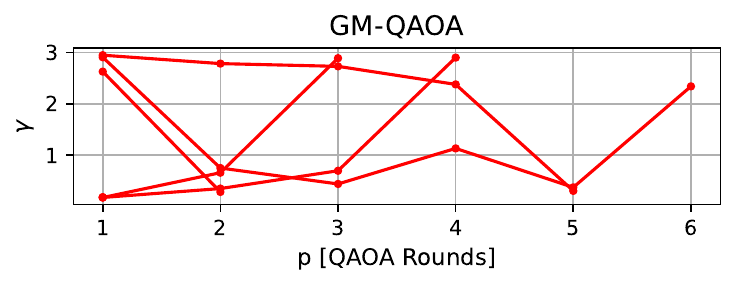}
    \caption{Learned QAOA angles from the Grover mixer QAOA simulations for the $3$ Ising models shown in Figure~\ref{fig:QAOA_fair_sampling_test_Ising_models}, up until convergence to the ground-state. The three vertical columns correspond to the three Ising models shown in Figure~\ref{fig:QAOA_fair_sampling_test_Ising_models}. For each step of $p$, the complete vectors of $\vec{\beta}$ (top row) and $\vec{\gamma}$ (bottom row) are shown. }
    \label{fig:QAOA_angles_GM}
\end{figure}

Figures~\ref{fig:QAOA_angles_X_mixer} and \ref{fig:QAOA_angles_GM} show the numeric values of the learned $\vec{\beta}$ and $\vec{\gamma}$ quantities used in Figure~\ref{fig:QAOA_fair_sampling_test_Ising_models}. 

\clearpage

\setlength\bibitemsep{0pt}
\printbibliography

\end{document}